\documentclass[%
 reprint,
nofootinbib,
 amsmath,amssymb,
 aps,
]{revtex4-2}

\usepackage[utf8]{inputenc}
\usepackage{dcolumn}% Align table columns on decimal point
\usepackage{bm}
\usepackage{physics}
\usepackage{bbold}
\usepackage{multirow}
\usepackage{makecell}
\PassOptionsToPackage{hyphens}{url}\usepackage[hidelinks]{hyperref}
\usepackage{graphics}
\usepackage{graphicx}

\newcommand*{\no}{\noindent}
\newcommand*{\bea}{\begin{eqnarray}}
\newcommand*{\eea}{\end{eqnarray}}
\newcommand*{\be}{\begin{equation}}
\newcommand*{\ee}{\end{equation}}

\newcommand*{\pref}[1]{(\ref{#1})}

\newcommand*{\nn}{\nonumber}

\newcommand{\bma}{\begin{pmatrix}}
\newcommand{\ema}{\end{pmatrix}}

\newcommand*{\la}{\left\langle}
\newcommand*{\ra}{\right\rangle}

\begin{document}

\title{Hints for a Geon from Causal Dynamic Triangulations}

\author{Axel Maas}
  \email{axel.maas@uni-graz.at}
\affiliation{
Institute of Physics, NAWI Graz, University of Graz, Universitätsplatz 5, 8010 Graz, Austria
}
\author{Simon Pl\"atzer}
 \email{simon.plaetzer@uni-graz.at}
\affiliation{
Institute of Physics, NAWI Graz, University of Graz, Universitätsplatz 5, 8010 Graz, Austria
}
\affiliation{Particle Physics, Faculty of Physics, University of Vienna, Boltzmanngasse 5, A-1090 Wien, Austria}

\author{Felix Pressler}
 \email{f.pressler@edu.uni-graz.at}
\affiliation{
Institute of Physics, NAWI Graz, University of Graz, Universitätsplatz 5, 8010 Graz, Austria
}

\date{\today}

\begin{abstract}
The existence of geons, physical states of self-bound gravitons, has long been proposed. In the context of four-dimensional causal dynamical triangulation simulations we investigate this possibility by measuring curvature-curvature correlators of different gravitational operators. We find a behavior consistent with a massive state, independent of the operators considered, over a certain distance window. While at most a hint, this is tantalizing due to its possible implications for dark matter or (primordial) black holes. We also find indications that the phase of rapid expansion of the obtained de Sitter universe impacts the mass, and relates to quantum fluctuations of space-time.
\end{abstract}

\maketitle

\section{Introduction} \label{sec:introduction}

In a loose sense, gravity can be considered to be a gauge theory of translation \cite{Kibble:1961ba,Sciama:1962,Hehl:1976kj}. The underlying gauge field, no matter whether this is taken to be the metric or something else, can thus not be directly observable. Physical observations require gauge invariant quantities derived from the metric, the simplest case being curvature scalars \cite{Wald:1984rg}. This holds in any  quantum (field) theory of gravity \cite{Ambjorn:2012jv,Maas:2019eux}, where physical observables need to be described in terms of correlation functions of such invariant quantities. Particles formed purely from gravitational degrees of freedom have been dubbed geons \cite{Wheeler:1955zz,Perry:1998hh} and should manifest within such gauge invariant correlation functions. Stable geons could constitute dark matter or even be (primordial) black holes.

The observables we consider, however, may not need to resemble particles at all, as the necessary notion of an asymptotic state of quantum field theory will in general not exist \cite{Parker:2009uva}. But, observation tells us that the notion of particle should at least be a good effective concept for distance regimes between billions of light years down to $10^{-3}$ fm. Hence, at least for some sets of parameters of quantum gravity and some distance scale, an effective particle picture makes sense.

Investigating these questions within quantum gravity is the aim of this work. While the questions raised apply to any version of quantum gravity, it may be answered differently in each.   In this work we exploit (causal) dynamical triangulations (CDT) \cite{Ambjorn:2012jv,Loll:2019rdj} which are accessible to non-perturbative simulations, feature a ``long-distance`` de Sitter-like dynamics \cite{Ambjorn:2012jv,Loll:2019rdj}, appear to be connected to asymptotic safety \cite{Ambjorn:2024bud} as another candidate for quantum gravity \cite{Niedermaier:2006wt,Reuter:2012id,Reuter:2019byg}, and allow to access both the curvature \cite{Klitgaard:2017ebu,Loll:2023hen} and correlation functions \cite{Dai:2021fqb,vanderDuin:2024pxb}. In particular the latter two are necessary ingredients for analyzing curvature correlators for signs of particle-like aspects.

We present our simulation setup in Sec.~\ref{s:tech}. In Sec.~\ref{s:geon} we discuss the definition of curvature correlators and the definition of a geon. Influenced by \cite{Maas:2019eux,Maas:2022lxv} and motivated by \cite{Dai:2021fqb}, we construct correlation functions which are designed to be able to resemble the ordinary behavior of correlation functions of massive particles in flat-space lattice field theory. While this can be well considered as an ad hoc procedure, the results in section \ref{s:results} surprisingly follow the expected pattern well. In fact, what we find indeed shows -- over an intermediate distance -- the behavior expected for a massive particle, which does not depend on details of the choice of operators. A brief appendix presents an analysis of some systematic uncertainties.

Of course, our results here are truly exploratory, and many questions, both conceptually and systematically, remain open. We will address them in future work. But, following here the seminal work \cite{vanderDuin:2024pxb}, we consider this to be a first step in establishing a similar phenomenology as we can muster in flat-space lattice field theory. And that is necessary for quantum gravity phenomenology using non-perturbative lattice-like methods.

\section{Technical setup}\label{s:tech}

As with ordinary lattice simulations, CDT simulations generate configurations, in terms of space-times triangulated by simplices, to approximate a path integral \cite{Ambjorn:2012jv,Loll:2019rdj}. On these configurations correlation functions are measured, and then statistically analyzed.

For the creation of the configurations we use the code of \cite{Ambjorn:2021yvk}. For technical details of CDT simulations and notations, we refer to \cite{Ambjorn:2012jv}. For the exploratory simulations at hand, we employed a single simulation point at the bare parameters $\Delta=0.6$ and $\kappa_0=2.2$, which exhibits a de Sitter-like space time on average \cite{Loll:2019rdj}, and thus a (classical) positive cosmological constant. Moreover, the lattice spacing, i.\ e.\ the (spatial) length $a$ of the simplices used to triangulate the space-time, has been estimated to be about $\approx2.1$ Planck lengths \cite{Ambjorn:2008wc}, which we will take here at face value. We performed simulations on $(N_t,N_\text{simp})=(60,80$k$)$, $(80,160$k$)$, and $(80,320$k$)$, with $N_t$ the temporal extent of the foliated space-time ({\it i.e.} the number of time slices) and $N_\text{simp}$ the number of (4,1)-simplices, which connect the time slices. Our setups are periodic in temporal direction and have spatially spherical boundary conditions. We note that the effective obtained geometry is found \cite{Ambjorn:2008wc,Klitgaard:2020rtv} to be four-dimensional spherical, and thus (Euclidean) de Sitter. Since we did not find any statistical significant dependence on the volume for our primary observables, i.\ e.\ within $2-3\sigma$, we will show in the following only results from the largest volume. Some more details on the volume dependence are presented in the appendix.

We denote quantities at fixed foliation a time-slice, and everything on or in between two neighboring foliations a fat time-slice. In total, we calculated for each case 5000 (2000 for $N_\text{simp}=320$k and 500 for the study of the $\delta$-dependence in the appendix) configurations, following standard procedures for thermalization and decorrelation \cite{Brunekreef:2023ljt}. 

In order to extract the correlation functions we discuss in Sec.~\ref{s:geon}, three primary quantities are needed per configuration: One is the number of simplices per time-slice, which is directly accessible. The second are spatial distances, which require more careful evaluation, and the third quantity are curvature scalars.

In order to extract the spatial distances, we could directly use the simplices' spatial edges with a fixed distance. However, it turns out that this is not the best spatial discretization. Following \cite{vanderDuin:2024pxb} we employ a dual triangulation. For that, the center of neighboring simplices are used to define a discrete distance element. Though the unique association with a fixed time-slice is lost in this case, the pseudo-time-like distance between two consecutive time-slices is expected to become infinitesimal in the continuum limit \cite{Ambjorn:2012jv,Loll:2019rdj}. We therefore neglect this effect rather than introducing a fractional time as was done in \cite{Ambjorn:2011ph,Klitgaard:2020rtv}. Instead, we consider all centers of simplices in what we call a ``fat'' foliation slice to be at the same effective foliation count. While we do not see any obvious effects related to this approximation, this will be resolved in more detail in future work.

Distance measures between non-neighboring points themselves are now non-trivial. Following \cite{Ambjorn:2012jv,Schaden:2015wia,Maas:2019eux}, we take as a space-like distance measure the geodesic distance $d$ between dual vertices on a fixed fat foliation slice. This is obtained by employing a Dijkstra algorithm, which is run on the triangulation restricted to a certain fat foliation slice. Since the determination using all possible pairs is, of course, prohibitively expensive, we have sampled 100 (50 for $N_\text{simp}=80$k) pairs per fixed fat foliation slice for a stochastic estimator \cite{Gattringer:2010zz}. Here, each vertex is individually sampled uniformly on the set of all vertices, and their distance determined afterwards. Thus, the distances are not uniformly sampled. This is corrected for by the normalization of the correlator with the correlator of the unity operator below, which is sampled in the same way. This is similar to the discussion in \cite{vanderDuin:2024pxb}, in particular their remarks in Appendix B, which we also took into account. We find that the average largest geodesic distance found increases like $\sim($number of pairs$)^{0.1}$, and thus very slowly. It reaches about 50\% of the largest found geodesic distance, which saturates quickly, at 100 pairs. This is, however, roughly also the largest relevant distance for the present purpose.

The final ingredient to build our operators is the Quantum Ricci Curvature Scalar (QRCS) $q$ \cite{Loll:2019rdj,Klitgaard:2017ebu,Loll:2023hen}. Given two points $p$ and $p'$, we can identify a sphere around each point of the pair by including all points with a distance $\delta$ from that point. If the geodesic distance between both sphere centers is $d(p,p')=\epsilon$, the normalized QRCS $Q$ is given by
\bea
Q(\epsilon)&=&\frac{q}{\delta}=\frac{1}{N(p,\delta)N(p',\delta)}\sum_{\substack{d(p,p')=\epsilon \\ d(q,p)=\delta,d(q',p')=\delta}} d(q,q')\nn\\
N(p,\delta)&=&\sum_{d(q,p)<\delta}1\nn,
\eea
\no where the sums run over all points of the dual triangulation, {\it i.e.} the centers of simplices. As $Q$ does no longer depend on $\epsilon$ explicitly but only through the summation, it is possible to identify $p$ and $p'$ by setting $\epsilon=0$, thus yielding a quantity measuring the distance between pairwise points on a sphere with respect to each other \cite{vanderDuin:2024pxb}. Thus, $Q$ is a measure of such distances, and triangulations with different local structure will in general yield locally different values of $Q$. In practice, we fixed $\delta=6$ for this work.  However, different values of $\delta$ should be considered to define different operators, akin to differently smeared operators in flat space-time \cite{Gattringer:2010zz}. Thus, long-distance physics should not be affected by the choice of $\delta$. We will give support that this is the case for our primary observables below and in the appendix. Other than that, the choice is here primarily technically motivated, as smaller $\delta$ have fewer points to average over the sphere, while larger ones become prohibitively expensive in terms of computer time.

If the theory has a smooth continuum limit to a Riemannian manifold then the limit $\delta\to 0$ yields \cite{Klitgaard:2018snm,vanderDuin:2024pxb}
\be
Q=q_1-\delta^2(q_3 R+{\cal O}(\delta))\label{ricci}
\ee
\no where the $q_i$ are positive, real constants, and $R$ is the standard, continuum, curvature scalar. Thus, $Q$ is in this limit a measure for curvature. Indeed, this was confirmed to be a good approximation of the functional dependency on $\delta$ in \cite{Klitgaard:2018snm} for CDT simulations. We confirmed this in the present work, if $\delta\gtrsim 4$. At smaller values of $\delta$ we did observe deviations, which are expected eventually when $\delta\approx a$. Below, we will give also a result for correlators build from $R$, which was obtained by fitting for each measurement of $Q$ to \pref{ricci} for $\delta\gtrsim 4$ to determine measurements of $R$.

Finally, we will be building one-point correlation functions, and (subtracted) two-point correlation functions \cite{vanderDuin:2024pxb,Ambjorn:1998vd}. As noted above, we uniformly sample vertices, and evaluate the operators at each of these vertices. For a one-point correlation function $\bar{O}=\la O\ra$ we average over all the locations per configuration on a fixed (fat) time slice, and then the result over all configurations. We also define fluctuation operators $\Delta O=O-\bar{O}$. Two-point correlation functions are calculated as a function of the geodesic distance $s$ between the point pair, by evaluating for each configuration
\be
\sum_{xy}\delta_{d(x,y),s}O(x)O(y)\label{cconf}
\ee
\no The operator $O$ in this case can also be a fluctuation operator $\Delta O$. Note that the number of points $x$ and $y$ will increase with distance for anything beyond an effectively one-dimensional system. Thus, the function will increase on average when increasing the number of point pair samples. This is not treated here, but will be taken care of as normalization in section \ref{s:geon}. The per-configuration correlator \pref{cconf} will then be averaged over all configurations we generate. Note that in general distances can exist at which no pairs of points is available. This is especially true for a stochastic sample. In such cases, the per-configuration correlator vanishes for this value of $s$.

\section{Defining a geon correlation function}\label{s:geon}

It is well established \cite{Loll:2019rdj,Klitgaard:2020rtv} that a single configuration, and even the average, in CDT simulations at the given simulation point do not resemble a flat Minkowski manifold, but are frequently close to a(n Euclidean) de Sitter space. This implies that it is impossible to speak about locations across configurations. In addition, the number of time-slices is usually not very large compared to the number of time-slices where the de Sitter space is almost static. It is a priori not clear, how the global space-time characteristics would affect the propagation of a gravitational state. We therefore opt for another approach.

In flat space-time states are (exponentially) screened in spatial directions. For a free particle in an isotropic space this screening mass would be the same as the pole mass. Since the CDT results do not suggest very strong gravitational interactions \cite{Ambjorn:2012jv,Loll:2019rdj}, it appears in this exploratory study justifiable to use the spatial screening mass as a proxy. This leaves the problem of comparing the same spatial structure across different configurations. However, the path integral of CDT at the simulation point is dominated by configurations, which individually are already close to a de Sitter universe \cite{Ambjorn:2012jv,Loll:2019rdj}, which we observe from our simulations, as well.

Thus, we define a cosmological time $\tau$, by setting it to zero at the (fat) foliation slice, where the spatial volume, approximated by the number of simplices, is maximal. All our measurements will be at fixed cosmological time relative to this origin, per configuration. Thus, the cosmological time is comparable between configurations. We then investigate spatial correlation functions
\be
C_{OO}(\tau,s)= \la \sum_{xy} \delta_{d(x,y),s} O(\tau,x) O(\tau,y)\ra\label{corr1}.
\ee
\no For the operators $O$ we choose $1$, $Q$, $Q^2$, and $R$, as defined in section \ref{s:tech}, where the unit operator effectively measures volume fluctuations. For all but the unit operator we also employ the subtracted versions, which are created in the same way as in \cite{vanderDuin:2024pxb,Ambjorn:1998vd}, in particular as $O\to O-\bar{O}=\Delta O$.

Finally, as noted already in section \ref{s:tech}, \pref{corr1} averages radially, and thus the number of suitable points increases with $s$. This increase is affected implicitly by the availability of points at a given distance, which in turn is affected by the structure of the manifold, and in an integral is giving rise to a factor $\sqrt{-\det g}$ of the metric. The first requires to normalize with the volume, the second with covariant volume. Following \cite{vanderDuin:2024pxb} we do this by building a normalized correlator
\be
D_{OO}(\tau,s)=\frac{C_{OO}(\tau,s)}{C_{11}(\tau,s)}\nn,
\ee
\no ensuring $D_{11}=1$. On any classical manifold, $C_{11}$ could be calculated exactly, but here it is a quantum observable itself, quantifying the correlation of volume fluctuations \cite{vanderDuin:2024pxb}. Moreover, by determining $C_{11}$ from the same pair sampling algorithm as $C_{OO}$, this will also correct for any non-uniformity in the sampling in the stochastic estimator.

Thus, $D_{OO}$ for $O=Q$ is a curvature-curvature correlator. Such a correlator, viewed as a correlator of gauge invariant interpolating operators for a geon, could indeed carry particle information \cite{Maas:2019eux,Maas:2022lxv}. If we are in a suitable distance window, we would expect that these function shows the same behavior as propagators in a flat-space quantum field theory, especially if the curvature is not too strong, i.\ e.\ a more or less exponential decay. Furthermore, as $Q^2$ has the same quantum numbers, an invariant scalar, a quasi-asymptotic state should appear in the same way also in the correlator $D_{Q^2Q^2}$, reflecting the fact that the choice of interpolating operator is arbitrary in the construction of the Lehmann-Symanczyk-Zimmermann (LSZ) reduction.

\section{Results}\label{s:results}

\begin{figure}
 \includegraphics[width=\columnwidth]{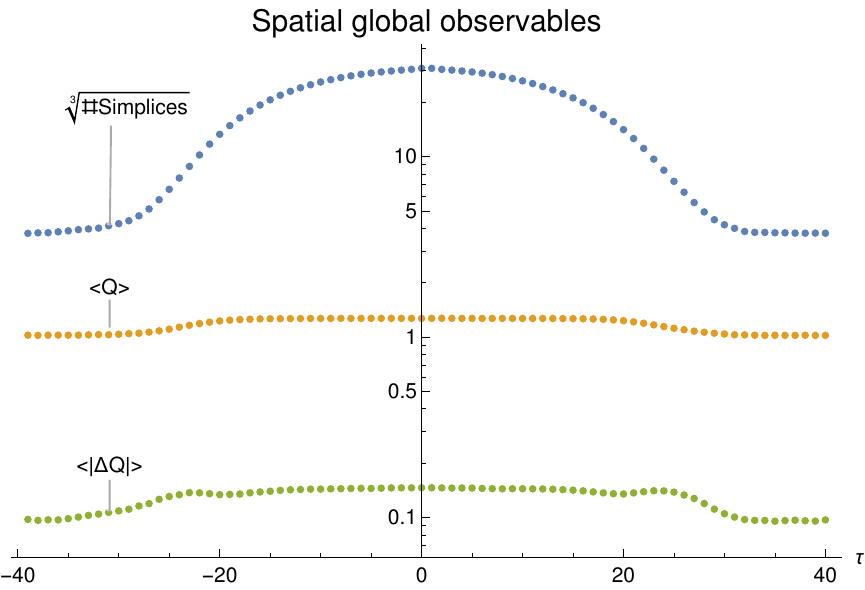}
 \caption{\label{global}Global properties as a function of cosmological time. The extent in simplices is measured by the cube root of all simplices with a vertex on a given fat time slice. Error bars are smaller than the plot points. }
\end{figure}

\begin{figure}
 \includegraphics[width=\columnwidth]{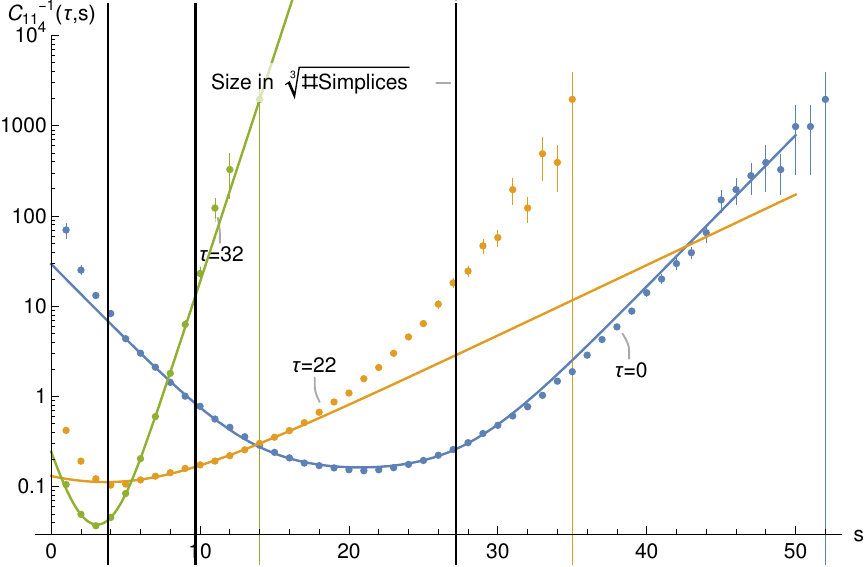}
 \caption{\label{i11}The inverse correlation function for $O=1$ with fits of type $a+b\cosh(m(s-s_0))$. It is shown for three cosmological times, which can be compared to figure \ref{global}: $\tau=0$ (blue) is the largest extent, $\tau=22$ (orange) is the inflection point of the $Q$ change, and $\tau=32$ (red) is the point where the system settles on the smallest extent. The perpendicular lines give the size at the corresponding $\tau$ in terms of the cube root of simplices, for illustration of the relative sizes.}
\end{figure}

\begin{figure}
 \includegraphics[width=\columnwidth]{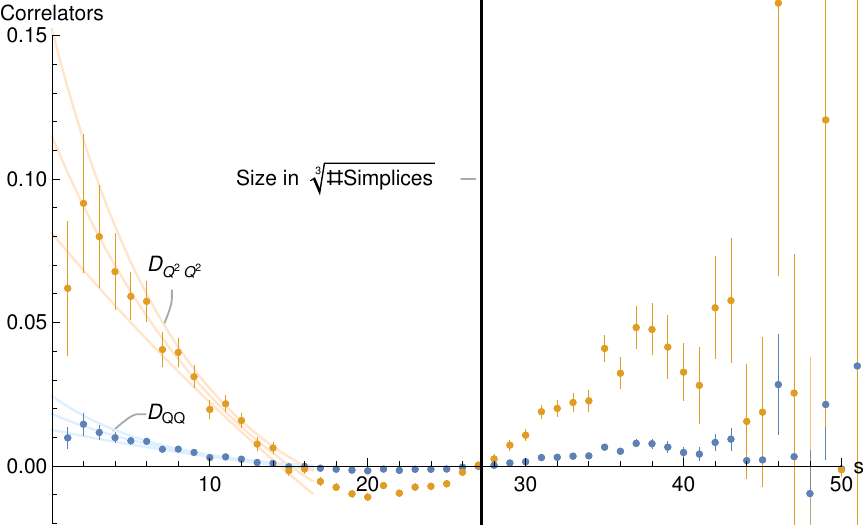}\\
 \includegraphics[width=\columnwidth]{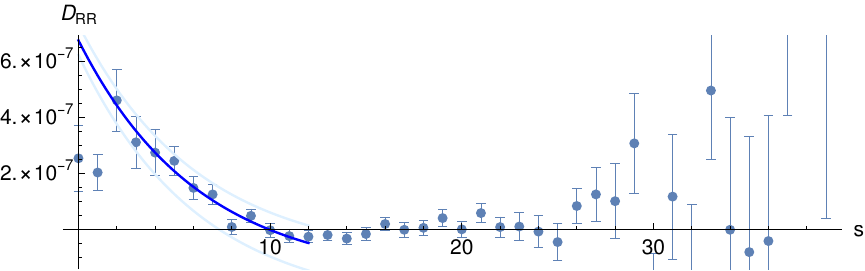}
 \caption{\label{prop}The normalized correlation function for $O=\Delta Q$ and  $O=\Delta Q^2$ with fits of type $a+b\exp(-ms)$ and their $\pm1\sigma$ error band (top panel) at $\tau=0$. The same for $O=R$, extracted using \pref{ricci} for each measurement of $Q$ (bottom panel), but for the smallest system. The fit in the lower panel fixes the same mass $m$ as in the top panel, and only determines $a$ and $b$. We note that the first two points in distance could be affected by discretization errors, and therefore deviate from an exponential.}
\end{figure}

The global properties are shown in figure \ref{global}. As is visible there is very little fluctuations in terms of the spatial size, which is here given in terms of the cube root of simplices per fat time slice\footnote{If the universe would be spatially perfectly spherical, the maximum geodesic distance is around half the sphere, and about 3 times larger. In the figures \ref{i11} and \ref{prop} this is visible, as the maximal spatial geodesic distance plotted there is around twice as large.}. Using \cite{Ambjorn:2008wc}, this implies a diameter of about 70 Planck lengths as maximum physical extent, and about 10 Planck lengths as minimal physical extent.

As noted previously  \cite{Ambjorn:2012jv,Loll:2019rdj}, the generated universes are of very similar properties, allowing to indeed identify our configuration-wise cosmological time to play a meaningful global role. We also observe the characteristic transition of the (Euclidean) de Sitter universe from a very small object to a very large one, which in Minkowski signature would be an exponential (inflationary) phase. We note that the average size of the QRCS changes, from a smaller value in the small-extent time to a larger value. This change is marked by a non-monotonous change in the curvature fluctuation, which peaks briefly close to the onset of the expansion. Surprisingly, fluctuations are largest not in the small extent time period, but rather in the large volume period.

We show the results for the inverse of the correlation function $C_{11}$, used for normalization, for three exemplary cosmological times in figure \ref{i11}, at maximum extent, at the fluctuation peak, and at minimum extent. The correlation function can be well fitted by an {\it Ansatz}
\be
a+b\cosh(m(s-s_0))\nn
\ee
\no with a constant term and a $\cosh$-behavior, with a parameter $s_0$ smaller than the maximal extent measured by the number of simplices, and a characteristic, $\tau-$dependent scale $m$, in dimensionless units. At $\tau=0$ its value is $\sim 0.39\to 0.27 M_{\text{Planck}}$, where the conversion has been done with the simplex constant 2.1 \cite{Ambjorn:2008wc}. 

With this, the subtracted and normalized correlation function for $O=\Delta Q$ and $O=\Delta Q^2$ can be constructed. They are shown exemplary at cosmological time $\tau=0$ in figure \ref{prop}. Initially, up to a distance of about $s\gtrsim15$, both can be fitted with the ansatz
\be
a+b\exp(-ms)\label{fitform}
\ee
\no fairly well, as shown in the figure. There are also no statistically significant short-distance artifacts visible, as have been seen in the two-dimensional case \cite{vanderDuin:2024pxb}, except perhaps at $s=1$. There is a negative offset $a$, but since this is not flat space, the usual arguments do not hold which would prevent such an effect. After that, both correlators again increase, until they eventually drown in noise before reaching the maximal possible spatial distances. Note that this distance is the maximal geodesic distance, and thus different from the estimate in simplices, which is smaller. In the same figure \ref{prop}, we also show the result for the correlation function of the curvature scalar $R$, obtained for each measurement of $Q$ separately using \pref{ricci} for values of $\delta\gtrsim 4$ on the smallest system. While significantly statistically more noisy, it shows the same behavior, and can be fitted with the same mass. This suggests that the result for the mass is not too dependent on $\delta$. This is in line with the fact, discussed in the appendix, that the $QQ$ correlator becomes fairly independent of $\delta$ above $\delta\gtrsim 4$. This gives confidence that the extracted mass parameter is a genuine physical quantity. A similar investigation on a larger system will, however, require more than an order of magnitude more computing time.

\begin{figure}
 \includegraphics[width=\columnwidth]{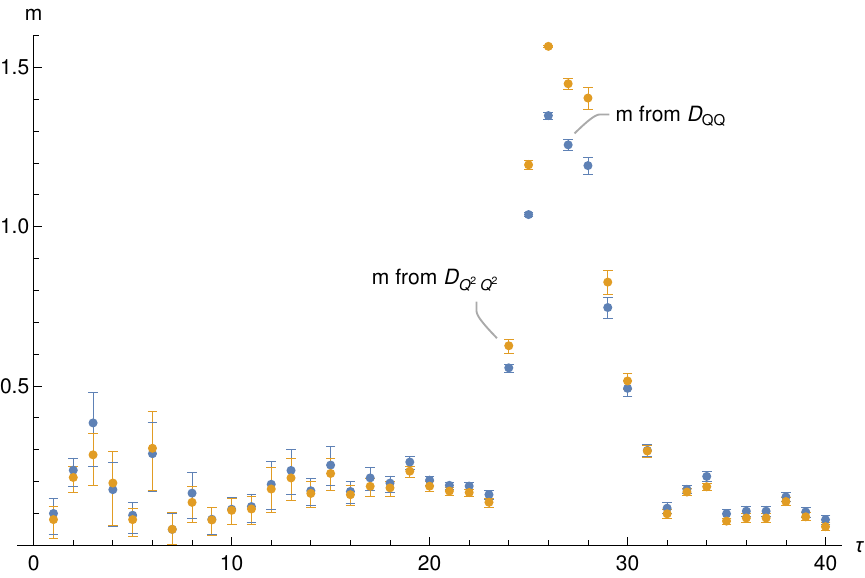}
 \caption{\label{mass}The extracted mass from the normalized correlators of $O=\Delta Q$ and $O=\Delta Q^2$.}
\end{figure}

Most importantly, the correlators for both operator choices $O=\Delta Q$ and $O=\Delta Q^2$, as well as $O=R$, display the same exponential decay, with a mass parameter which agrees within errors. The extracted mass values $m$ from the fit \pref{fitform} is shown for the choices $O=\Delta Q$ and $O=\Delta Q^2$ in Fig.~\ref{mass}. That they agree within error is what is expected if the result is independent of the choice of interpolating operators in the LSZ reduction. The result is thus, over sufficiently short distances $s\lesssim 40l_\text{Planck}\sim l_\text{Universe}/2$, consistent with a massive particle-like behavior, and thus a geon interpretation. The fact that the behavior changes over longer distances is worth noting and raises questions about possible explanations.

The shape of the correlators remains roughly the same, and gets only more compressed with increasing cosmological time, and thus decreasing spatial extent of the universe. However, as Fig.~\ref{mass} shows, the mass parameter, determining the exponential decay, is relatively stable, except during the phase of rapid expansion, where it quickly rises. It remains, however, approximately the same for both operators. This suggest that during the expansion of the Universe in our CDT simulations, the properties of geons changes drastically, but are more or less constant at times of (roughly) constant extension. This pattern is the same for all volumes, and we observe no statistically significant volume dependence. The masses, averaged over $0\le\tau<12$, are found to be 0.18(1), 0.16(1), 0.17(3) for the $Q$ correlator and 0.14(1), 0.17(1), 0.15(3) for the $Q^2$ correlator, for $N_\text{simp}$ being 80k, 160k, and 320k, respectively. It would correspond to a mass of $\sim 0.09 M_\text{Planck}$ using the result of \cite{Ambjorn:2008wc}.

We note that our system is relatively coarse, and thus the conversion to physical units may well have associated discretization artifacts. Only investigations along lines-of-constant physics can eventually clarify such a potential dependence. However, the fact that the result is within a few $\sigma$ independent on the maximal spatial extent suggests that the mass will eventually not be zero. Thus, the number should be considered for illustrative purposes only at the present time.

\section{Summary}\label{s:sum}

Our results are consistent with the already observed global features of CDT simulations \cite{Ambjorn:2012jv,Loll:2019rdj}. Once we rearrange our measurements relative to the extension of the universe, we do see strong changes with the cosmological time, and especially during the phase of rapid expansion. But we always see that over short and intermediate distances we have a particle-like behavior of our curvature-curvature correlators, which is the same for all three operators employed, as required for an effective LSZ construction. Thus, the interpretation of self-bound gravitons as geons is tempting. More investigations will be needed to figure out whether this is accurate, and, if yes, if these objects are more like a potential dark matter candidate, a primordial black hole, or something else entirely. At any rate, taken at face value, a geon appears possible with a mass of order Planck mass, and also being a sensitive probe for the phases of rapid expansion. Alone the possibility of such radically different physical objects is tantalizing.\\

\noindent{\bf Acknowledgments} \\

We are grateful to D\'aniel N\'emeth for support, and him and all of the authors of \cite{Ambjorn:2021yvk} for providing us with the code from \cite{Ambjorn:2021yvk}. We are grateful to him and Renate Loll for discussions. We are grateful to Renate Loll, D\'aniel N\'emeth, and Jesse van der Duin for useful comments on the manuscript. Simulations have been performed on the GSC at Graz University.

\section*{Appendix: Systematic effects}

\section*{Smearing}

\begin{figure}
 \includegraphics[width=\columnwidth]{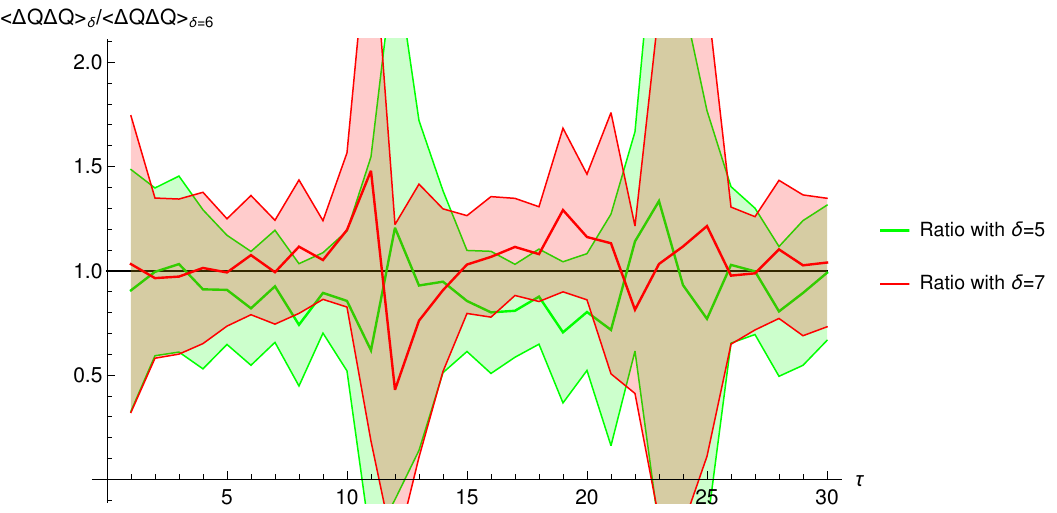}\\
 \includegraphics[width=\columnwidth]{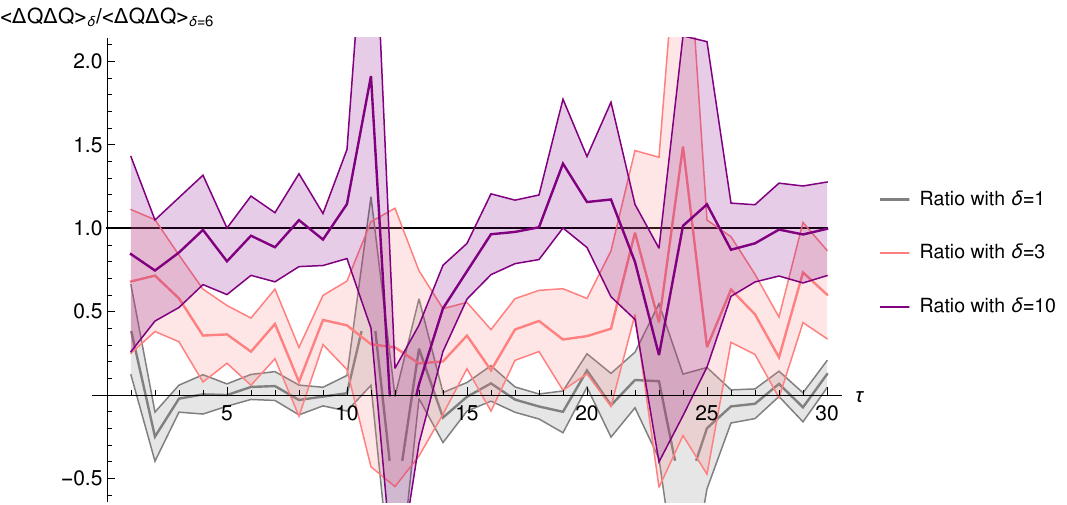}
 \caption{\label{dratio}The ratio of the correlation function of $\Delta Q$ for different values of $\delta$ to the one at $\delta=6$. The top panel shows the situation for $\delta=5$ and $\delta=7$, i.\ e.\ around the value of $\delta=6$ used primarily in the text. The lower panel shows more extreme cases of $\delta=1$, $\delta=3$, and $\delta=10$. The bands give the $1\sigma$ error band calculated from error propagation. Results are from the smallest system $(60,80$k$)$.}
\end{figure}

To ensure that the parameter $\delta$ is indeed only acting as a smearing parameter, without distorting the physics, a twofold strategy can be performed. One, the probably more accurate one, is to extract the curvature scalar using \pref{ricci}, as was done in the main part of the text and shown in figure \ref{prop}. However, this comes at a substantial penalty in terms of statistics. In this appendix, as an alternative, we study the correlator of $\Delta Q$ as a function of $\delta$. To this end, the ratio of the correlator at different values of $\delta$  to the correlator at $\delta=6$, the one primarily used in the main text, is shown in figure \ref{dratio}. In this way the normalization drops out.

It is visible that if $\delta=1$, and thus the edge length of the simplices, the ratio is very different from one. This is to be expected, as there the smearing is sampling mainly lattice artifacts. Even at $\delta=3$, the ratio is different from one, but starts to get flatter. For $5\le \delta\le 7$, the ratio is basically flat, and the correlator essentially independent, within errors, of $\delta$, as expected from a smearing parameter. Note that the uncertainties grow naturally very large at the zero crossings. Even at $\delta=10$, where the smearing radius starts to become comparable to half the spatial extent of the universes, the ratio remains mostly flat.

Thus, as argued in the main text, the interpretation of $\delta$ as a smearing parameter and the choice of $\delta=6$ is well justified, within the available systematics. This is in line with the observations from the extracted curvature scalar, which should be independent of $\delta$.

\section*{Volume}

\begin{figure}
 \includegraphics[width=\columnwidth]{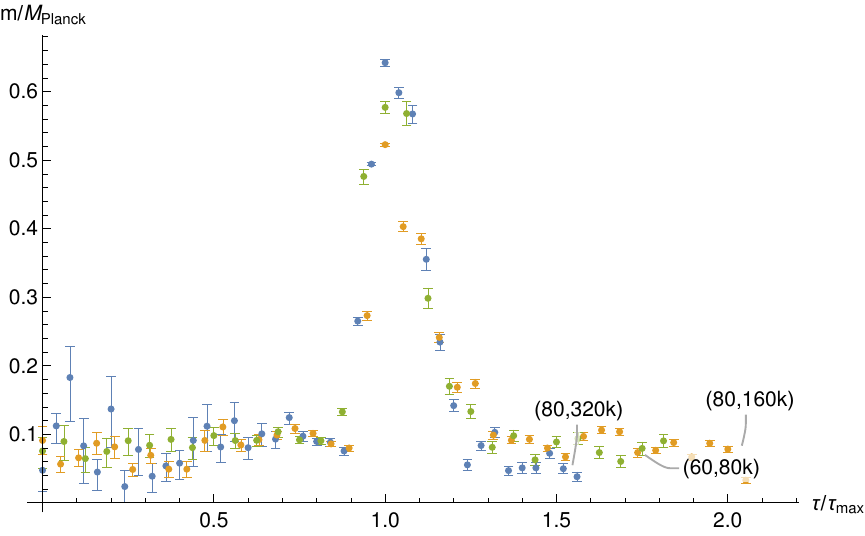}
 \caption{\label{massv} The dependency on the system size of the extracted mass. The cosmological time has been rescaled to the cosmological time at which the mass peaks, $\tau_\text{max}$. This figure was first shown in \cite{Maas:2025mne}.}
\end{figure}

To understand the volume dependence, the cosmological time-dependence of the extracted mass of the correlator is shown for different volumes (and temporal extents) in figure \ref{massv}. Since the obtained systems have a characteristic scale, the cosmological time extent of the large-volume behavior, the cosmological time is here rescaled to the time $\tau_\text{max}$. This time is where the extracted mass has its largest value which, as noted in the main text, roughly coincides with the quick changes of spatial extent. As is visible, after this rescaling the value and the cosmological time extension dependence coincides in all systems. The only feature where a dependency is seen is the maximal mass at $\tau_\text{max}$, which, however, is not a critical feature for the present purpose. Thus, as stated in the main text, the critical part of the behavior of the correlator is fairly independent on the size of the system in simplices or the temporal extent.

\bibliographystyle{bibstyle}
\bibliography{bib}

\end{document}